\documentclass[aps, pra, showpacs, superscriptaddress, twocolumn, 10pt]{revtex4-1}

\usepackage{amsfonts,amssymb,amsmath,amsthm,longtable,array}
\usepackage{graphics,graphicx,color,enumitem}
\usepackage{epstopdf,epsfig}

\def\identity{\leavevmode\hbox{\small1\kern-3.8pt\normalsize1}}
\def\openone{\leavevmode\hbox{\small1 \normalsize \kern-.64em1}}
\newcommand{\ba}{\begin{align}}
\newcommand{\ea}{\end{align}}
\newcommand{\bpf}{\begin{proof}}
\newcommand{\epf}{\end{proof}}
\newcommand{\ket}[1]{ | #1 \rangle}
\newcommand{\bra}[1]{ \langle #1  |}

\newcommand{\proj}[1]{\ket{#1}\bra{#1}}

\DeclareMathAlphabet{\mathpzc}{OT1}{pzc}{m}{it}
\newcommand{\pv}{\mathpzc{p}_V}
\newcommand{\Pv}{\mathpzc{P}_V}
\newcommand{\Plr}{\mathpzc{P}_{LR}}

\newcounter{rowNo}
\begin{document}
\title{Multipartite nonlocality and random measurements}
\author{Anna~de~Rosier}
\affiliation{Institute of Theoretical Physics and Astrophysics, Faculty of Mathematics, Physics and Informatics, University of Gda\'nsk, 80-308 Gda\'nsk, Poland}
\author{Jacek~Gruca}
\affiliation{Institute of Theoretical Physics and Astrophysics, Faculty of Mathematics, Physics and Informatics, University of Gda\'nsk, 80-308 Gda\'nsk, Poland}
\author{Fernando~Parisio}
\affiliation{Departamento de F\'isica, Federal University of Pernambuco, Recife, PE 50670-901, Brazil}
\author{Tam\'as~V\'ertesi}
\affiliation{Institute for Nuclear Research, Hungarian Academy of Sciences, H-4001 Debrecen, P.O. Box 51, Hungary}
\author{Wies{\l}aw~Laskowski}
\affiliation{Institute of Theoretical Physics and Astrophysics, Faculty of Mathematics, Physics and Informatics, University of Gda\'nsk, 80-308 Gda\'nsk, Poland}

\begin{abstract}
We present an exhaustive numerical analysis of violations of local realism by families of multipartite quantum states. As an indicator of nonclassicality we employ the probability of violation for randomly sampled observables. Surprisingly, it rapidly increases with the number of parties or settings and even for relatively small values local realism is violated for almost all observables. We have observed this effect to be typical in the sense that it emerged for all investigated states including some with randomly drawn coefficients. We also present the probability of violation as a witness of genuine multipartite entanglement.
\end{abstract}

\maketitle

\section{Introduction}
Quantum multiparticle systems do not provide a mere amplification of the nontrivial effects displayed by two-party systems.
Rather, they bring about completely new phenomena and applications.  On the fundamental level, multipartite systems, e. g., have been employed to illustrate nonlocality without Bell inequalities \cite{GHZ} and, more recently, to show that finite-speed superluminal causal influences would allow for superluminal signalling between spatially separated parties \cite{gisinNP}. In what concerns applications, one-way quantum computing \cite{oneway} and multipartite secret sharing \cite{liang-gisin} are outstanding examples where complex quantum systems can be employed.

As is the case for multipartite entanglement, the characterization of nonclassical features of multiparticle systems is a hard problem with
several open questions \cite{bellNL}. One interesting possibility to analyze the nonclassicality of complex states is to study their correlation properties under random measurements. With this motivation we will be concerned with the following quantity
\begin{equation}
\Pv(\rho)=\int f (\Omega)d\Omega,
\label{prob}
\end{equation}
where the integration variables correspond to all parameters that can be varied within a Bell scenario and, $f=1$ only for settings that lead to violations in local realism, and vanishes otherwise. Note that, when properly normalized, $\Pv$ can be interpreted as a probability of violation of local realism.

The probability $\Pv$ can be used at different context levels. One can select a particular Bell inequality $I$ and integrate $f_I$ over
all possible settings of the corresponding Bell experiment. This was mainly the approach adopted in previous theoretical \cite{PhysRevLett.104.050401,PhysRevA.83.022110} and experimental \cite{Liang} works.
This is also the case of ref. \cite{PhysRevA.92.030101}, where the quantity defined in (\ref{prob}) has been considered as a measure of nonlocality and applied in the context of the Collins-Gisin-Linden-Massar-Popescu (CGLMP) inequality \cite{CGLMP,PhysRevA.65.052325}. This procedure, however, would face increasing difficulties as the number of parties grows.
For a relatively modest number of qubits, e. g., the corresponding number of inequivalent Bell inequalities with a fixed (say 2) number of settings is already very large and, thus, addressing one inequality at a time would become prohibitive.
On a deeper level we can dispense with the choice of a particular inequality and directly consider the space of behaviors (space of joint probabilities), which local polytopes inhabit. In this case, the integration refers to all possible measurements, the only context information required being the number of measurements per party. This is the approach that we will adopt here, so that we use the probability of violation to evaluate the degree of nonclassicality of several relevant states involving up to five qubits and also bipartite states of qutrits.

This work is presented in the following way. In the next section we provide a brief description of the numeric method to be employed (linear programming).
In section III we present our results in the form of several tables and discuss their main consequences. In the last section we give our final remarks and some perspectives.

\section{Description of the method}
In our numerical analysis we consider the most general Bell experiment with $N$ spatially separated observers performing measurements on a given state of $N$ qu$d$its with $d=2$ (qubits) and $d=3$ (qutrits). Each observer can choose among $m_i$ arbitrary observables $\{O^i_{1}, O^i_{2}, ..., O^i_{m_i}\}$ ($i=1,2,...,N$) defined by orthogonal projections $O^i_{j} = \sum_{r_i=0}^{d-1} r^i \proj{v^i_j}$ linked by the general unitary transformations $\ket{v^i_j} = U^i_j |r^i \rangle$. The unitary transformations are parametrized by three angles for qubits:
\begin{equation}
U^i_{j} (\phi_1^{i,j},\psi_1^{i,j},\chi_1^{i,j})=
\left(\begin{array}{c c}
\cos\phi_1^{i,j}\text{ e}^{i\psi_1^{i,j}} & \sin\phi_1^{i,j} \text{ e}^{i\chi_1^{i,j}} \\
-\sin\phi_1^{i,j}\text{ e}^{-i\chi_1^{i,j}} & \cos\phi_1^{i,j} \text{ e}^{-i\psi_1^{i,j}}
\end{array} \right),
\end{equation}
and eight angles for qutrits:
\begin{eqnarray}
U^i_{j}& (\phi_1^{i,j},\psi_1^{i,j},\chi_1^{i,j},\phi_2^{i,j},\psi_2^{i,j},\chi_2^{i,j},\phi_3^{i,j},\psi_3^{i,j})=\nonumber\\
&\left(\begin{array}{c c c}
	\cos\phi_1^{i,j}\text{e}^{i\psi_1^{i,j}} & \sin\phi_1^{i,j} e^{i\chi_1^{i,j}} & 0 \\
	-\sin\phi_1^{i,j} \text{e}^{-i\chi_1^{i,j}} & \cos\phi_1^{i,j} \text{e}^{-i\psi_1^{i,j}} & 0 \\
	0 & 0 & 1
\end{array} \right)\nonumber\\
&\times \left(\begin{array}{ccc}
	\cos\phi_2^{i,j}\text{e}^{i\psi_2^{i,j}} & 0 & \sin\phi_2^{i,j} \text{e}^{i\chi_2^{i,j}}  \\
	0 & 1 & 0\\
	-\sin\phi_2^{i,j} \text{e}^{-i\chi_2^{i,j}} & 0 & \cos\phi_2^{i,j} \text{e}^{-i\psi_2^{i,j}} \\
\end{array} \right) \nonumber\\
&\times \left(\begin{array}{ccc}
	1 & 0 & 0\\
	0 & \cos\phi_3^{i,j}\text{e}^{i\psi_3^{i,j}} & \sin\phi_3^{i,j}  \\
	0 & -\sin\phi_3^{i,j} & \cos\phi_3^{i,j} \text{e}^{-i\psi_3^{i,j}} \\
\end{array} \right).
\end{eqnarray}

A local realistic description of an experiment is equivalent to the existence of a joint probability distribution
$p_{\rm lr}(r_1^1, ..., r_{m_1}^1, ..., r_1^N, ..., r_{m_N}^N)$,
where $r_{j_i}^i = \{0,1,...,d-1\}$ denotes the result of the measurement of the $i$th observer's $O^i_{j}$ observable.
If the model exists, quantum predictions for the probabilities are given by the marginal sums:
\begin{eqnarray}
&P(r^1, ..., r^N \left| O^1_{k_1}, ..., O^N_{k_N}   \right.) & \nonumber \\ 
&={\rm Tr}(\rho \proj{v^1_{k_1}} \otimes \cdots \otimes \proj{v^N_{k_N}})& \label{set-eq} \\ 
&=\sum\limits_{r^1_{j_1}, ..., r^N_{j_N}=0}^{d-1} p_{\rm lr}(r^1_1, ..., r_{m_1}^1, ..., r_1^N, ..., r_{m_N}^N)& \nonumber
\end{eqnarray}
where $P(r^1, ..., r^N \left|O^1_{k_1}, ..., O^N_{k_N}\right.)$ denotes the probability of obtaining the result $r^i$ by the $i$th observer while measuring observables $O^i_{k_i}$ and $j_i \neq k_i$ ($i=1,..., N$).
It can be shown that for some quantum entangled states the marginal sums cannot be satisfied, which is an expression of Bell's theorem.

Our task is to find, for a given state $\rho$ and a set of observables $O^i_{k_i}$ ($i=1,...,N$; $k_i = 1, ..., m_i)$,
whether the local realistic model exists, i.e.,  all the equations (\ref{set-eq}) can be satisfied.
This can be done by means of linear programming (see e.g. \cite{PhysRevLett.85.4418, PhysRevA.82.012118, j.cam.2013.12.003}). It is worth mentioning that the method allows us to reveal nonclassicality even without direct knowledge of Bell inequalities for the given experimental situation.

Finally, we check how many sets of settings (in percents) lead to violation of local realism. We introduce a frequency $\pv(\rho)$ which for a sufficiently large statistics converges to the probability of violation $\Pv(\rho)$. We provided sufficient statistics to not observe changes in results on the third decimal place.

The measurement operators are sampled according to Haar measure \cite{Haar}.
The angles $\psi_r$ and $\chi_r$ are taken from uniform distributions on the intervals:
$ 0\leq\psi_r <2\pi$ and $0\leq\chi_r<2\pi.$
To generate $\phi_r$ in interval $0\leq\phi_r\leq\frac{\pi}{2}$ it is convenient to use an auxiliary random variable $\xi_r$ distributed uniformly on $0\leq\xi_r<1$ and $\phi_r=\arcsin(\xi_r^{1/2})$ for $r\in\left\{1,2\right\}$ and $\phi_3=\arcsin(\xi_r^{1/4})$. Of course, all variables are generated independently for each observer $i$ and measurement setting $j$.

\section{Results and analysis}

We applied the numerical method to prominent families of quantum states:
\begin{itemize}
\item[(1)] the generalized $N$ qubit GHZ state \cite{GHZ,GHZ.VS.NG}
\begin{equation}
\ket{\rm{GHZ}(\alpha)}_N=\sin \alpha \ket{0...0}_N+\cos \alpha \ket{1...1}_N, \nonumber
\end{equation}
and for $\alpha = \pi/4$, $\ket{\rm{GHZ}(\alpha)}_N \equiv \ket{\rm{GHZ}}_N$;
\item[(2)] the four qubit singlet state \cite{singlet4,PhysRevA.68.012304}
\begin{eqnarray}
\ket{\psi_4^-}&=&\frac{1}{\sqrt{3}}\left(\ket{0011}+\ket{1100}\right)\nonumber \\
&-&\frac{1}{\sqrt{12}}\left(\ket{0101}+\ket{0110}+\ket{1001}+\ket{1010}\right); \nonumber
\end{eqnarray}
\item[(3)] the $N$ qubit Dicke state with $e$ excitations \cite{Dicke}
\begin{equation}
\ket{D_N^e}=\binom{N}{e}^{-1/2}\sum_{\rm permutations}|0...0\underbrace{1...1}_{e}0...0\rangle_N, \nonumber
\end{equation}
where the special case $e=1$ is referred to as the $N$ qubit W state $\ket{W}_N\equiv\ket{\rm{D}_N^1}$ \cite{W};
\item[(4)] the four qubit cluster state \cite{cluster}
\begin{equation}
\ket{\rm{Cluster}_4}=\frac{1}{2}\left(\ket{0000}+\ket{0011}+\ket{1100}-\ket{1111} \right); \nonumber
\end{equation}
\item[(5)] the generalized $N$ qutrit GHZ state
\begin{equation}
\ket{\rm{GHZ}^{d=3}(\alpha)}_N=\sin\alpha\ket{0...0} +\frac{1}{\sqrt{2}}\cos\alpha(\ket{1...1}+\ket{2...2});\nonumber
\end{equation}
\item[(6)] the three qutrit singlet state (Aharonov state) \cite{Aharonov}
\begin{eqnarray}
\ket{A^-}_3&=&\frac{1}{\sqrt{6}}(\ket{012}+\ket{120}+\ket{201} \nonumber\\
           &-&\ket{011}-\ket{101}-\ket{110});\nonumber
\end{eqnarray}
\item[(9)] the three qutrit Dicke states with the sum of excitations equals to $e$ \cite{DickeQt}
\begin{eqnarray}
\ket{\rm{Q}^1_3}&=&\frac{1}{\sqrt{3}}\sum_{\pi} \pi\left\{\ket{001}\right\}, \nonumber\\
\ket{\rm{Q}^2_3}&=&\frac{1}{\sqrt{15}}\left(2\sum_{\pi}\pi\left\{\ket{011}\right\}+\sum_{\pi}\pi\left\{\ket{002}\right\}\right),\nonumber\\
\ket{\rm{Q}^3_3}&=&\frac{1}{\sqrt{10}}\left(2\ket{111}+\sum_{\pi}\pi\left\{\ket{012}\right\}\right).\nonumber
\end{eqnarray}
\end{itemize}

We calculated the frequencies $\pv(\rho)$ for an increasing number of different settings per site. All results are presented in Tables \ref{tab-qubits},  \ref{tab-random} and \ref{tab-qutrits}. Some states which appear on the tables are not listed above. They will be defined in the appropriate paragraphs. Our results lead to the following observations.

\subsubsection{Comparison with known results}
The probability of violation was previously examined in several contexts.
The only analytical result on tight inequalities was obtained in \cite{PhysRevLett.104.050401} for the simplest scenario of two settings and two outcomes, where the probability of violation of different versions of the CHSH inequality \cite{CHSH} has been obtained by the two qubit GHZ state (the Bell state). In this case our numerical method gives the same value  (No.~\ref{GHZ2}) as the analytical expression
$\pv(\rm{GHZ}_2)=\Pv^{\rm CHSH}(\rm{GHZ}_2) =2(\pi-3) \sim 0.283183$ with accuracy to four decimal places.

For $N>2$, the GHZ state has been studied only numerically. In \cite{PhysRevLett.104.050401} the state was analyzed in the context of WWW{\.Z}B inequality for $N \leq 6$. In \cite{PhysRevA.83.022110} the analysis was extended to $N=15$ qubits (WWW{\.Z}B inequality) and $N=6$ (using a similar linear programming method). In all cases, the results agree with our numerical method.

\subsubsection{Genuine tripartite entanglement criterion}

We note that for any two-qubit state and two measurement settings per party, the probability of violation of local realism cannot be greater than $2(\pi-3)$, i.e., the two-qubit GHZ state gives the highest probability. The analytical proof is deferred to the Appendix.

Then, it is straightforward to prove that for any biproduct state $\ket{\psi_{12}}\otimes\ket{\psi_3}$ the two-qubit quantum probability $P\left(r_1,r_2\left|A_i,B_j\right.\right)$ is described by a local realistic theory if and only if $P\left(r_1,r_2,r_3\left|A_i,B_j,C_k\right.\right)$ does. Hence, in the examined cases of entangled states of $N_E$ particles, multiplied by the product state $\ket{0}^{\otimes N_0}$,  the full $(N_E+N_0)$-particle state has, as expected, exactly the same probability of violation as its entangled component alone. The above property comes along with the fact that biseparable states (i.e. convex mixtures of biproduct states) can only lower the probability of violation compared to biproduct states. So we can argue that for any 3-qubit state (including mixed states) with two measurement settings per party, if $\Pv(\rho)>2(\pi-3)$, this certifies that the three qubit state is genuinely tripartite entangled, that is, it can not be written in any of the forms $\ket{\psi_{12}}\otimes\ket{0}$, $\ket{\psi_{13}}\otimes\ket{0}$ and $\ket{0}\otimes\ket{\psi_{23}}$ and convex combinations of these states.
Indeed, data in the table \ref{tab-qubits} indicates that both GHZ3 and W3 states are genuinely tripartite entangled as the respective probabilities: 74.688\%  (No.~\ref{GHZ3}) and 54.893\%  (No.~\ref{W3}) are much higher than 28.319\%.

One could construct a similar condition for higher number of parties ($N>3$) but in this case one may give only numerical bounds for the critical probability, because analytical results are not known in these cases.

We also considered the probability of violation for the state $\psi_3(\theta)= \cos\theta\ket{111}+\sin\theta\ket{\rm{W}_3}$  (Nos.~\ref{PSI15}-\ref{PSI90}). For all values of angle $\theta>25.975^{\circ}$ one can prove that the state is genuinely three-partite entangled \cite{PhysRevLett.88.170405}, whereas our numerical method reveals the threshold slightly below $30^\circ$.
This discrepancy, though small, is due to the fact that our criterion is a necessary but not a sufficient one.

\subsubsection{Non-additivity and multiplicative features of $\Pv(\rho)$}
The question of additivity seems to be better posed in terms of $\Pv(\rho)$ than in terms of maximal violations of a Bell inequality. Consider the example of the state $\ket{\rm{GHZ}_2}\otimes\ket{\rm{GHZ}_2}$, for which probability of violation is non-additive, since $\pv(\ket{\rm{GHZ}_2}\otimes \ket{\rm{GHZ}_2})\approx 1.7 \pv(\ket{\rm{GHZ}_2})$, and is a bit less than half of $\pv(\ket{\rm{GHZ}_4})$.

Therefore instead of additivity, we should consider the multiplicative features of $\Pv(\rho)$.
Concerning $\pv(\ket{\rm{GHZ}_2})$ and $\pv(\ket{\rm{GHZ}_2}\otimes \ket{\rm{GHZ}_2})$, the probabilities that measurement results admit a local realistic description, $\Plr=1-\Pv$, should be multiplied.
In this particular case,
\begin{eqnarray}
&&\Plr(\ket{\rm{GHZ}_2}) = 1 - \Pv(\ket{\rm{GHZ}_2}) = 1-2(\pi-3), \\
&&\Plr(\ket{\rm{GHZ}_2} \otimes \ket{\rm{GHZ}_2}) = \Plr(\ket{\rm{GHZ}_2})^2 = (7-2\pi)^2. \nonumber
\end{eqnarray}
Hence, $\pv(\rho_{\ket{\rm{GHZ}_2}}\otimes \rho_{\ket{\rm{GHZ}_2}}) = 1-(7-2\pi)^2 = 0.486176$
which fits our numerical results up to displayed digits  (No.~\ref{GHZ2GHZ2}).

We also examined the product of the two qubit GHZ state with a state that does not violate any two setting Bell inequality, namely the Werner state: $\rho_{\rm{Werner}_2}  = 1/\sqrt{2} |{\rm GHZ}\rangle_2 \langle {\rm GHZ}| + (1-1/\sqrt{2})\openone/4.$ In this case the probability of violation for the resulting state is the same as for $\ket{\rm{GHZ}_2}$, what can be explained by the above multiplicative feature, since $\Plr(\rho_{\rm{Werner}_2})=1$.

\subsubsection{Non-maximal probability of violation for GHZ states of more than 3 particles}
We observe a surprising feature, which emerges if the number of qubits is larger than three. It is well known that the $N$-qubit GHZ state maximizes many entanglement conditions and measures \cite{RevModPhys.81.865}. However, already for $N=4$ the probability of violation for the cluster state (No. \ref{cluster2222}) is greater than for the GHZ state (No. \ref{GHZ4}). The situation is even more dramatic for $N=5$, where the probability is greater for any out of 10 randomly sampled pure states  (Nos. \ref{rand5a}-\ref{rand5j}).

There is a particular entanglement measure which is in pace with the above observations, namely the generalized Schmidt Rank (SR) \cite{cluster}, corresponding to the minimal number of product states required to represent a given state. The SR of a GHZ state is two for any number of qubits, and it has been shown in \cite{cluster} that the SR behaves as $2^{\left\lfloor N\slash 2\right\rfloor}$ for cluster states of $N$ qubits.

\subsubsection{All typical states of five or more qubits violate local realism for almost all settings}

Even with only two observables per party it becomes almost impossible not to detect non-classicality for states with 5 qubits or more.
Any of the studied states (Nos.~\ref{GHZ5}-\ref{R5}) including random 5-qubit states (Nos. \ref{rand5a}-\ref{rand5j}) lead to nearly 100$\%$ probability of violation. In fact, the numbers are so close, that one can not distinguish the states by means of the violation probability. This amounts to an enhancement of the content of Gisin's theorem in the sense that not only all entangled states seem to be nonclassical, but they violate local realism for almost all experimental situations. That is, given an entangled state it is very likely that one can prove its non-classicality on a first try by choosing random observables (note also related recent results in Ref.~\cite{sampling}). This is to be contrasted with the original demonstration \cite{GISIN1991201,GISIN199215}, involving two qubits, where the settings have to be carefully selected. Of course, one can always find some states with a $\pv(\rho)$ which is much smaller than 100\%  (e.g. Nos. \ref{32}, \ref{23}), but they are strictly less entangled.

\subsubsection{$\pv(\rho)$ rapidly increases with the number of settings}

The probability of violation increases significantly also with the number of settings per party. For the two qubit GHZ state, and five measurement settings per site, the corresponding violation probability is almost equal to 1. This means that almost all randomly sampled settings lead to a conflict with local realistic models and to the violation of some Bell inequality.

This rapid growth is more pronounced than it is for robustness against white noise admixture. An increase is also observed in the resistance to noise, but it is usually a much less evident effect and visible particularly in multipartite cases \cite{PhysRevA.82.012118}. For example, due to the recent work~\cite{brierley2016convex}, an increase of $0.58\%$ in the noise resistance of the two-qubit maximally entangled state required 30 settings (see also a previous work~\cite{VERTESI}). It is also conjectured that the above improvement in the noise resistance could not be attained with fewer settings. Note also that one cannot go beyond the increase of $3.682\%$ in noise resistance using an infinite number of projective measurements~\cite{hirsch2016better}.

\begin{figure}
	\centering
		\includegraphics[width=0.48\textwidth]{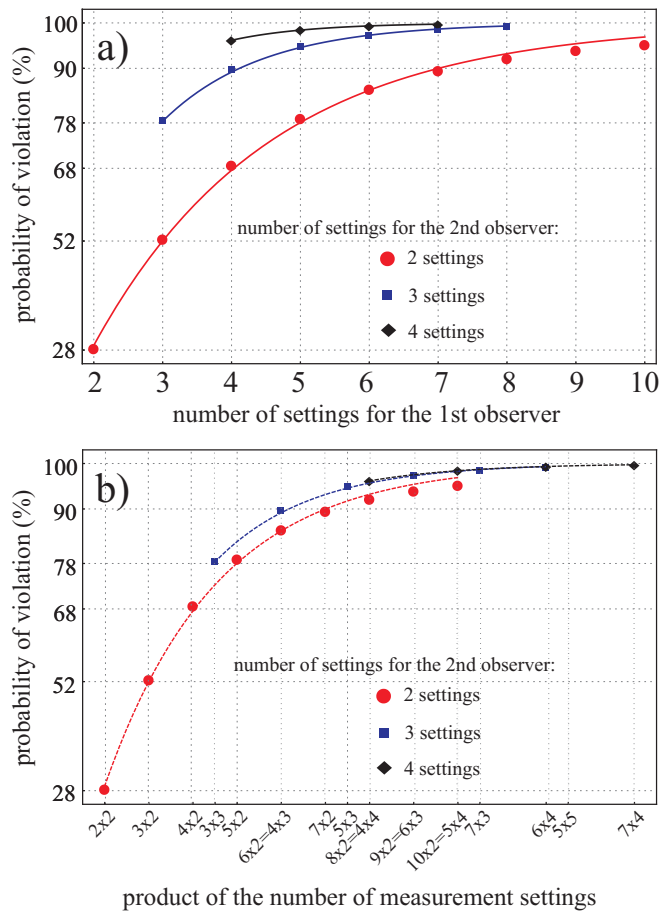}
		\caption{\label{fig-sett}The probability of violation for the two qubit GHZ states vs. (a) the number of measurement settings for the first observer; (b)a product of the number of settings for both observers.}
\end{figure}

The dependence of $\pv(\rho)$ as a function of the number of settings can be approximated by $1 - a e^{-bx}$, where $a,b$ are constant parameters and $x$ can be either the number of settings referring to one party (with the other number of settings fixed) (Fig. \ref{fig-sett}a) or a product of the number of possible measurement settings (Fig. \ref{fig-sett}b). Of course there are other possible combinations involving the number of settings.

\subsubsection{Nonclassicality of bound entangled states}

A bound entangled state (BES) is entangled but undistillable \cite{HOR}. However, in \cite{PhysRevA.74.010305} it was shown that the 4-qubit bound entangled Smolin state \cite{PhysRevA.63.032306} can maximally violate a 2-setting Bell inequality similar to the standard CHSH inequality. In accordance with this finding, when numerically investigating the possibility of a local realistic description for the Smolin state, even with 2 settings per party, we get small, but nonzero probability of violation, $\pv(\rho_{\rm{Smolin}_4})$=0.023\%  (No.~\ref{Smolin2222}). Although this value is three orders of magnitude smaller than that for other examined entangled states, it grows very fast (faster than for other entangled states) with the number of settings  (Nos.~\ref{Smolin3222}-\ref{Smolin3333}) and the growth seems to be exactly exponential.

In general, if we investigate the $\Pv$ of a PPT state \cite{peres1996,horodecki1996} and find it to be non-vanishing, then the state must be entangled. Note that this conclusion can be reached even without the knowledge of which Bell inequality is to be violated. This may be particularly useful when the state involves many subsystems.
In general, if we investigate the $\Pv$ of a PPT state \cite{peres1996,horodecki1996} and find it to be non-vanishing, then the state must be entangled. Note that this conclusion can be reached even without the knowledge of which Bell inequality is to be violated. This may be particularly useful when the state involves many subsystems.

The three qubit BES $\rho_{\rm BES}^{2\times 2\times 2}$ introduced in \cite{VB12} also violates some Bell inequality, but this seems to be statistically very rare since we had not observed any violation of local realism for two settings per party. Nevertheless, when the same measurement is applied to every particle, we observed a nonzero probability of violation, $\pv(\rho_{\rm BES}^{2\times 2\times 2})$=0.008\%.

The last considered example of bound entangled states is the two qutrit state $\rho_{\rm BES}^{3\times 3}$ that was used to disprove the famous Peres conjecture \cite{peres1999all,ncomms6297}.
Despite the fact that this state does not admit a local realistic model, the violation is proved only for judiciously specified observables and inequality, which occur seldom enough, so that we did not find violations in any of the $10^{10}$ randomly chosen settings.

\subsubsection{Two qutrits: coincidence of maximal entanglement and maximal nonclassicality}
\begin{figure}
\centering
\includegraphics[width=0.48\textwidth]{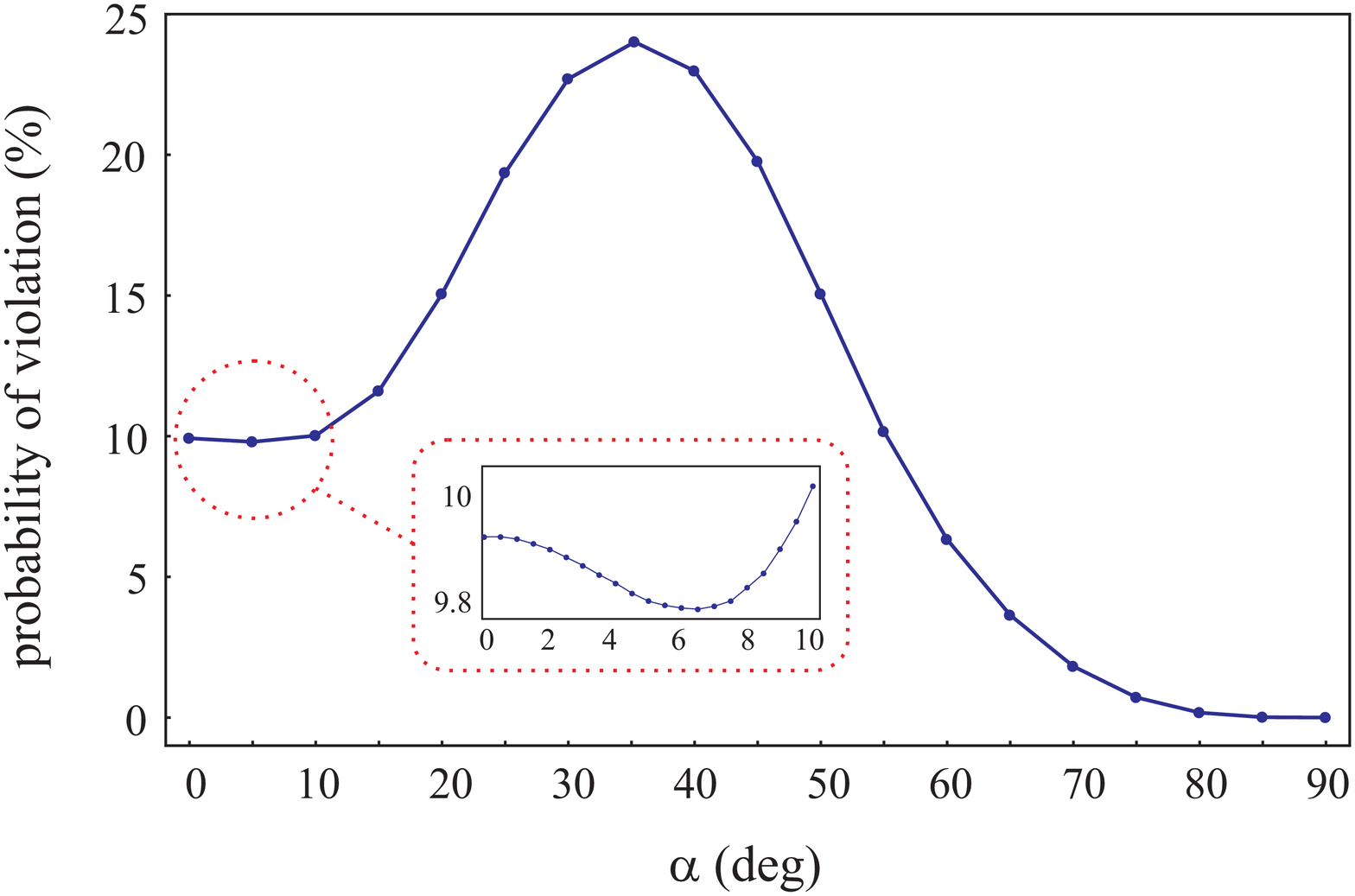}
\caption{\label{fig-qutrit} Probability of violation for 2 qutrit generalized GHZ($\alpha$) state vs. $\alpha$}
\end{figure}

Entanglement and nonclassicality are distinct resources.
The former corresponds to the purely mathematical concept of state nonseparability while the later amounts to its manifestation in experiments. It is acknowledged that a clear illustration of this point is the unexpected difference between maximally entangled states and states that maximally violate a Bell inequality.
In \cite{PhysRevA.92.030101} it is suggested that this anomaly may be an artifact of almost all measures that have been used to quantify nonclassicality.
Our numerical results show that, according to the probability of violation, there is no anomaly in the nonclassicality of two qutrit generalized GHZ state. The maximal probability of violation $\pv$=24.011\% (No.~\ref{symmGHZqt2}) is attained for the symmetric state $\rm{GHZ}^{d=3}_2(35.26^\circ)$ instead of the asymmetric one: $\rm{GHZ}^{d=3}_2(29.24^\circ)$ [$\pv=22.317\%$ (No.~\ref{asymmGHZqt2})], which maximally violates the CGLMP inequality \cite{PhysRevA.65.052325}. A little surprising is the behavior of the probability of violation around $\alpha =0$, where we observe a small local minimum for $\alpha = 6^\circ$ (see Fig. \ref{fig-qutrit}). The minimum remains even if the number of settings per party is increased to three. A possible explanation of this feature could be the fact that there are two relevant Bell inequalities for the considered case -- CHSH and CGLMP inequalities with different functions representing the violation probability. The total probability of violation is a combination of the probabilities for those particular inequalities, what may result in several extremes.

\begin{footnotesize}
\begin{longtable}{>{\the\numexpr\value{rowNo}}>{\refstepcounter{rowNo}}cccccc}
\caption{\label{tab-qubits} Frequencies of violation of local realism $\pv$ observed statistically with random measurements on qubit states}
\endfirsthead \endhead
\hline \hline
\multicolumn{1}{c}{No.} & $N$ & State & Settings & Stat. & $\pv$ \\
\hline\hline
\label{GHZ2} & 2 & $\ket{\rm{GHZ}_2}$ & $2\times2$ & $10^{10}$ & 28.318 \\ %
& 2 & $\ket{\rm{GHZ}_2}$ & $3\times2$ & $10^9$ & 52.401 \\
& 2 & $\ket{\rm{GHZ}_2}$ & $4\times2$ & $10^9$ & 68.654 \\
& 2 & $\ket{\rm{GHZ}_2}$ & $5\times2$ & $10^9$ & 78.947 \\
& 2 & $\ket{\rm{GHZ}_2}$ & $6\times2$ & $10^9$ & 85.391 \\
& 2 & $\ket{\rm{GHZ}_2}$ & $7\times2$ & $10^9$ & 89.482 \\
& 2 & $\ket{\rm{GHZ}_2}$ & $8\times2$ & $10^9$ & 92.150 \\
& 2 & $\ket{\rm{GHZ}_2}$ & $9\times2$ & $10^8$ & 93.945 \\
& 2 & $\ket{\rm{GHZ}_2}$ & $10\times2$& $10^8$ & 95.198 \\ \cline{4-6}
& 2 & $\ket{\rm{GHZ}_2}$ & $3\times3$ & $10^9$ & 78.219 \\
& 2 & $\ket{\rm{GHZ}_2}$ & $4\times3$ & $10^9$ & 89.545 \\
& 2 & $\ket{\rm{GHZ}_2}$ & $5\times3$ & $10^9$ & 94.658 \\
& 2 & $\ket{\rm{GHZ}_2}$ & $6\times3$ & $10^9$ & 97.085 \\
& 2 & $\ket{\rm{GHZ}_2}$ & $7\times3$ & $10^9$ & 98.303 \\
& 2 & $\ket{\rm{GHZ}_2}$ & $8\times3$ & $10^8$ & 98.953 \\ \cline{4-6}
& 2 & $\ket{\rm{GHZ}_2}$ & $4\times4$ & $10^9$ & 96.169 \\
& 2 & $\ket{\rm{GHZ}_2}$ & $5\times4$ & $10^8$ & 98.460 \\
& 2 & $\ket{\rm{GHZ}_2}$ & $6\times4$ & $10^8$ & 99.321 \\
& 2 & $\ket{\rm{GHZ}_2}$ & $7\times4$ & $10^8$ & 99.672 \\ \cline{4-6}
& 2 & $\ket{\rm{GHZ}_2}$ & $5\times5$ & $10^8$ & 99.504 \\
\cline{3-6} 
& 2 & $\ket{\rm{GHZ}_2(1^\circ)}$  & $2\times2$ & $10^{10}$ & 0.00000025\\
& 2 & $\ket{\rm{GHZ}_2(10^\circ)}$ & $2\times2$ & $10^9$ & 0.093 \\
& 2 & $\ket{\rm{GHZ}_2(20^\circ)}$ & $2\times2$ & $10^9$ & 2.826 \\
& 2 & $\ket{\rm{GHZ}_2(30^\circ)}$ & $2\times2$ & $10^9$ & 14.796 \\
& 2 & $\ket{\rm{GHZ}_2(40^\circ)}$ & $2\times2$ & $10^9$ & 26.599 \\
\cline{2-6} 
& 3 & $\ket{\rm{GHZ}_2}\otimes \ket{0}$ & $2\times2\times2$ & $10^9$ & 28.317 \\
& 3 & $\ket{\rm{GHZ}_2}\otimes \ket{0}$ & $3\times2\times2$ & $10^9$ & 52.399 \\
\cline{3-6} 
\label{GHZ3} & 3 & $\ket{\rm{GHZ}_3}$ & $2\times2\times2$ & $10^9$ & 74.688 \\
						 & 3 & $\ket{\rm{GHZ}_3}$ & $3\times2\times2$ & $10^9$ & 90.132 \\
						 & 3 & $\ket{\rm{GHZ}_3}$ & $4\times2\times2$ & $10^9$ & 95.357 \\
						 & 3 & $\ket{\rm{GHZ}_3}$ & $3\times3\times2$ & $10^9$ & 97.245 \\
						 & 3 & $\ket{\rm{GHZ}_3}$ & $4\times3\times2$ & $10^8$ & 98.926 \\ %
						 & 3 & $\ket{\rm{GHZ}_3}$ & $4\times4\times2$ & $10^8$ & 99.590 \\ %
						 & 3 & $\ket{\rm{GHZ}_3}$ & $3\times3\times3$ & $10^9$ & 99.542 \\
\cline{3-6} 
					 & 3 & $\ket{\rm{W}_3}$ & $1\times2\times2$ & $10^9$ & 15.244 \\
\label{W3} & 3 & $\ket{\rm{W}_3}$ & $2\times2\times2$ & $10^9$ & 54.893 \\
					 & 3 & $\ket{\rm{W}_3}$ & $3\times2\times2$ & $10^9$ & 76.788 \\
					 & 3 & $\ket{\rm{W}_3}$ & $4\times2\times2$ & $10^9$ & 87.287 \\
					 & 3 & $\ket{\rm{W}_3}$ & $5\times2\times2$ & $10^9$ & 92.465 \\
					 & 3 & $\ket{\rm{W}_3}$ & $3\times3\times2$ & $10^9$ & 91.366 \\
					 & 3 & $\ket{\rm{W}_3}$ & $3\times3\times3$ & $10^9$ & 97.797 \\
\cline{3-6} 
\label{PSI15} & 3 & $\ket{\rm{\psi}_3(15^\circ)}$ & $2\times2\times2$ & $10^9$ & 4.941 \\
							& 3 & $\ket{\rm{\psi}_3(20^\circ)}$ & $2\times2\times2$ & $10^9$ & 10.327 \\
							& 3 & $\ket{\rm{\psi}_3(25^\circ)}$ & $2\times2\times2$ & $10^8$ & 18.762 \\ %
							& 3 & $\ket{\rm{\psi}_3(25.975^\circ)}$ & $2\times2\times2$ & $10^9$ & 20.786 \\
							& 3 & $\ket{\rm{\psi}_3(30^\circ)}$ & $2\times2\times2$ & $10^9$ & 30.323 \\
							& 3 & $\ket{\rm{\psi}_3(45^\circ)}$ & $2\times2\times2$ & $10^9$ & 64.382 \\
							& 3 & $\ket{\rm{\psi}_3(60^\circ)}$ & $2\times2\times2$ & $10^9$ & 74.689 \\
							& 3 & $\ket{\rm{\psi}_3(75^\circ)}$ & $2\times2\times2$ & $10^9$ & 65.377 \\
\label{PSI90} & 3 & $\ket{\rm{\psi}_3(90^\circ)}$ & $2\times2\times2$ & $10^9$ & 54.893 \\
\cline{2-6} 
& 4 & $\ket{\rm{GHZ}_2}\otimes \ket{00}$ & $2\times2\times2\times2$  & $10^8$ & 28.318 \\
& 4 & $\ket{\rm{GHZ}_2}\otimes \ket{00}$ & $3\times2\times2\times2$  & $10^7$ & 52.407 \\
\cline{3-6}
\label{GHZ2GHZ2} & 4 & $\ket{\rm{GHZ}_2}\otimes \ket{\rm{GHZ}_2}$ & $2\times2\times2\times2$ & $10^8$ & 48.617 \\
								 & 4 & $\ket{\rm{GHZ}_2}\otimes \ket{\rm{GHZ}_2}$ & $3\times2\times2\times2$ & $10^7$ & 65.887 \\
\cline{3-6}
& 4 & $\ket{\rm{GHZ}_3}\otimes \ket{0}$ & $2\times2\times2\times2$ & $10^8$ & 74.683 \\
& 4 & $\ket{\rm{GHZ}_3}\otimes \ket{0}$ & $3\times2\times2\times2$ & $10^7$  & 90.134 \\
\cline{3-6}
\label{GHZ4} & 4 & $\ket{\rm{GHZ}_4}$ & $2\times2\times2\times2$ & $10^8$ & 94.240 \\
						 & 4 & $\ket{\rm{GHZ}_4}$ & $3\times2\times2\times2$ & $10^8$ & 98.352 \\
						 & 4 & $\ket{\rm{GHZ}_4}$ & $4\times2\times2\times2$ & $10^7$ & 99.339 \\ %
						 & 4 & $\ket{\rm{GHZ}_4}$ & $3\times3\times2\times2$ & $10^7$ & 99.624 \\
						 & 4 & $\ket{\rm{GHZ}_4}$ & $4\times3\times2\times2$ & $10^6$ & 99.867 \\
						 & 4 & $\ket{\rm{GHZ}_4}$ & $4\times4\times2\times2$ & $10^5$ & 99.937 \\
						 & 4 & $\ket{\rm{GHZ}_4}$ & $3\times3\times3\times2$ & $10^7$ & 99.934 \\
						 & 4 & $\ket{\rm{GHZ}_4}$ & $4\times3\times3\times2$ & $10^5$ & 99.981 \\
						 & 4 & $\ket{\rm{GHZ}_4}$ & $4\times4\times3\times2$ & $10^5$ & 99.989 \\
						 & 4 & $\ket{\rm{GHZ}_4}$ & $4\times4\times4\times2$ & $10^5$ & 99.993 \\
						 & 4 & $\ket{\rm{GHZ}_4}$ & $3\times3\times3\times3$ & $10^6$ & 99.995 \\
						 & 4 & $\ket{\rm{GHZ}_4}$ & $4\times3\times3\times3$ & $10^5$ & 99.999 \\
						 & 4 & $\ket{\rm{GHZ}_4}$ & $4\times4\times3\times3$ & $10^5$ & 99.999 \\
						 & 4 & $\ket{\rm{GHZ}_4}$ & $4\times4\times4\times3$ & $10^5$ & 100.00 \\
						 & 4 & $\ket{\rm{GHZ}_4}$ & $4\times4\times4\times4$ & $10^4$ & 100.00 \\ %
\cline{3-6} 
\label{W4} & 4 & $\ket{\rm{W}_4}$ & $2\times2\times2\times2$ & $10^8$ & 85.920 \\
					 & 4 & $\ket{\rm{W}_4}$ & $3\times2\times2\times2$ & $10^7$ & 95.129 \\
					 & 4 & $\ket{\rm{W}_4}$ & $4\times2\times2\times2$ & $10^7$ & 97.969 \\
					 & 4 & $\ket{\rm{W}_4}$ & $5\times2\times2\times2$ & $10^6$ & 99.013 \\ %
					 & 4 & $\ket{\rm{W}_4}$ & $3\times3\times2\times2$ & $10^7$ & 98.757 \\
					 & 4 & $\ket{\rm{W}_4}$ & $3\times3\times3\times2$ & $10^7$ & 99.767 \\
					 & 4 & $\ket{\rm{W}_4}$ & $3\times3\times3\times3$ & $10^6$ & 99.966 \\
					 & 4 & $\ket{\rm{W}_4}$ & $4\times4\times4\times2$ & $10^5$ & 99.999 \\
\cline{3-6} 
& 4 & $\ket{\rm{D}_4^2}$ & $2\times2\times2\times2$ & $10^8$ & 83.577 \\
& 4 & $\ket{\rm{D}_4^2}$ & $3\times2\times2\times2$ & $10^7$ & 94.065 \\
& 4 & $\ket{\rm{D}_4^2}$ & $4\times2\times2\times2$ & $10^7$ & 97.315 \\
& 4 & $\ket{\rm{D}_4^2}$ & $3\times3\times2\times2$ & $10^7$ & 98.428 \\
& 4 & $\ket{\rm{D}_4^2}$ & $3\times3\times3\times2$ & $10^7$ & 99.716 \\
& 4 & $\ket{\rm{D}_4^2}$ & $3\times3\times3\times3$ & $10^6$ & 99.964 \\
& 4 & $\ket{\rm{D}_4^2}$ & $4\times4\times4\times2$ & $10^5$ & 99.996 \\
\cline{3-6} 
& 4 & $\ket{\psi_4^-}$ & $2\times2\times2\times2$ & $10^8$ & 74.943 \\
& 4 & $\ket{\psi_4^-}$ & $3\times2\times2\times2$ & $10^7$ & 89.604 \\
& 4 & $\ket{\psi_4^-}$ & $4\times2\times2\times2$ & $10^7$ & 94.918 \\
& 4 & $\ket{\psi_4^-}$ & $3\times3\times2\times2$ & $10^7$ & 96.621 \\
& 4 & $\ket{\psi_4^-}$ & $3\times3\times3\times2$ & $10^7$ & 99.344 \\
& 4 & $\ket{\psi_4^-}$ & $3\times3\times3\times3$ & $10^6$ & 99.908 \\
& 4 & $\ket{\psi_4^-}$ & $4\times4\times4\times2$ & $10^5$ & 99.991 \\
\cline{3-6} 
\label{cluster2222} & 4 & $\ket{\rm{Cluster}_4}$ & $2\times2\times2\times2$ & $10^8$ & 97.283 \\
										& 4 & $\ket{\rm{Cluster}_4}$ & $3\times2\times2\times2$ & $10^8$ & 99.275 \\
										& 4 & $\ket{\rm{Cluster}_4}$ & $4\times2\times2\times2$ & $10^7$ & 99.705 \\
										& 4 & $\ket{\rm{Cluster}_4}$ & $3\times3\times2\times2$ & $10^7$ & 99.884 \\
										& 4 & $\ket{\rm{Cluster}_4}$ & $3\times3\times3\times2$ & $10^7$ & 99.976 \\
										& 4 & $\ket{\rm{Cluster}_4}$ & $3\times3\times3\times3$ & $10^6$ & 99.997 \\
										& 4 & $\ket{\rm{Cluster}_4}$ & $4\times4\times4\times2$ & $10^5$ & 99.999 \\
\cline{3-6} 
\label{Smolin2222} & 4 & $\rho_{\rm{Smolin}_4}$ & $2\times2\times2\times2$ & $10^8$ & 0.023 \\
\label{Smolin3222} & 4 & $\rho_{\rm{Smolin}_4}$ & $3\times2\times2\times2$ & $10^8$ & 0.068 \\
									 & 4 & $\rho_{\rm{Smolin}_4}$ & $4\times2\times2\times2$ & $10^7$ & 0.127 \\ %
									 & 4 & $\rho_{\rm{Smolin}_4}$ & $5\times2\times2\times2$ & $10^7$ & 0.195 \\ %
									 & 4 & $\rho_{\rm{Smolin}_4}$ & $3\times3\times2\times2$ & $10^7$ & 0.197 \\
									 & 4 & $\rho_{\rm{Smolin}_4}$ & $3\times3\times3\times2$ & $10^7$ & 0.601 \\
\label{Smolin3333} & 4 & $\rho_{\rm{Smolin}_4}$ & $3\times3\times3\times3$ & $10^7$ & 2.009 \\
\cline{2-6} 
\label{GHZ5} & 5 & $\ket{\rm{GHZ}_5}$ & $2\times2\times2\times2\times2$ & $10^7$ & 99.601 \\ %
& 5 & $\ket{\rm{GHZ}_5}$ & $3\times2\times2\times2\times2$ & $10^6$ & 99.900 \\ %
\cline{3-6} 
& 5 & $\ket{\rm{W}_5}$ & $2\times2\times2\times2\times2$ & $10^7$ & 98.311 \\ %
\cline{3-6} 
& 5 & $\ket{\rm{D}_5^2}$ & $2\times2\times2\times2\times2$ & $10^7$ & 99.254 \\ %
\cline{3-6} 
& 5 & $\frac{1}{2}(\rho_{\rm{D}_5^2}+\rho_{\rm{D}_5^3})$ & $2\times2\times2\times2\times2$ & $10^7$ & 0.047  \\
\cline{3-6} 
& 5 & $\ket{\rm GHZ}_4 \ket{0} $ & $2\times2\times2\times2\times2$ & $10^7$ & 94.240  \\
\cline{3-6}
\label{32}& 5 & $\ket{\rm GHZ}_3 \ket{00} $ & $2\times2\times2\times2\times2$ & $10^7$ & 74.688  \\
\cline{3-6}
\label{23}& 5 & $\ket{\rm GHZ}_2 \ket{000} $ & $2\times2\times2\times2\times2$ & $10^7$ & 28.318  \\
\cline{3-6}   
& 5 & $\ket{\rm{L}_5}$ \cite{PhysRevA.73.022303} & $2\times2\times2\times2\times2$ & $10^6$ & 99.782 \\
\cline{3-6} 
\label{R5} & 5 & $\ket{\rm{R}_5}$ \cite{PhysRevA.73.022303}& $2\times2\times2\times2\times2$ & $10^6$ & 99.957 \\
\hline \hline
\end{longtable}
\end{footnotesize}

\begin{footnotesize}
\begin{table}
\caption{\label{tab-random} Frequencies of violation of local realism $\pv$ observed statistically with random measurements on random qubit states}
\begin{tabular}{>{\the\numexpr\value{rowNo}}>{\refstepcounter{rowNo}}ccccc}
\hline \hline
\multicolumn{1}{c}{No.} & $N$ & Settings & Stat. & $\pv$   \\
\hline\hline
& 3 & $2\times2\times2$ & $10^8$ & 12.396 \\ 
& 3 & $2\times2\times2$ & $10^8$ & 33.893 \\ 
& 3 & $2\times2\times2$ & $10^8$ & 38.959 \\ 
& 3 & $2\times2\times2$ & $10^8$ & 45.186 \\ 
& 3 & $2\times2\times2$ & $10^8$ & 43.505 \\ 
& 3 & $2\times2\times2$ & $10^8$ &  4.812 \\ 
& 3 & $2\times2\times2$ & $10^8$ & 59.824 \\ 
& 3 & $2\times2\times2$ & $10^8$ & 35.197 \\ 
& 3 & $2\times2\times2$ & $10^9$ & 43.602 \\ 
& 3 & $2\times2\times2$ & $10^8$ & 43.747 \\ 
\cline{2-5}
& 4 & $2\times2\times2\times2$ & $10^7$ & 95.016 \\ 
& 4 & $2\times2\times2\times2$ & $10^7$ & 93.104 \\ 
& 4 & $2\times2\times2\times2$ & $10^7$ & 95.630 \\ 
& 4 & $2\times2\times2\times2$ & $10^7$ & 90.957 \\ 
& 4 & $2\times2\times2\times2$ & $10^7$ & 92.616 \\ 
\cline{2-5}
\label{rand5a}& 5 & $2\times2\times2\times2\times2$ & $10^7$ & 99.862 \\ 
& 5 & $2\times2\times2\times2\times2$ & $10^7$ & 99.857 \\ 
& 5 & $2\times2\times2\times2\times2$ & $10^7$ & 99.900 \\ 
& 5 & $2\times2\times2\times2\times2$ & $10^7$ & 99.889 \\ 
& 5 & $2\times2\times2\times2\times2$ & $10^7$ & 99.913 \\ 
& 5 & $2\times2\times2\times2\times2$ & $10^7$ & 99.878 \\ 
& 5 & $2\times2\times2\times2\times2$ & $10^7$ & 99.884 \\ 
& 5 & $2\times2\times2\times2\times2$ & $10^7$ & 99.880 \\ 
& 5 & $2\times2\times2\times2\times2$ & $10^7$ & 99.861 \\ 
\label{rand5j}& 5 & $2\times2\times2\times2\times2$ & $10^7$ & 99.878 \\ 
\hline \hline
\end{tabular}
\end{table}
\end{footnotesize}

\begin{footnotesize}
\begin{longtable}{>{\the\numexpr\value{rowNo}}>{\refstepcounter{rowNo}}cccccc}
\caption{\label{tab-qutrits} Frequencies of violation of local realism $\pv$ observed statistically with random measurements on qutrit states}
\endfirsthead \endhead
\hline \hline
\multicolumn{1}{c}{No.} & $N$ & State & Settings & Stat. & $\pv$ \\
\hline\hline
& 2 & $\ket{\rm{GHZ}^{d=3}(0^\circ)}_2$  & $2\times2$ & $10^9$ & 9.925 \\
& 2 & $\ket{\rm{GHZ}^{d=3}(5^\circ)}_2$  & $2\times2$ & $10^9$ & 9.801 \\
& 2 & $\ket{\rm{GHZ}^{d=3}(10^\circ)}_2$ & $2\times2$ & $10^8$ & 10.021 \\
& 2 & $\ket{\rm{GHZ}^{d=3}(15^\circ)}_2$ & $2\times2$ & $10^7$ & 11.609 \\
& 2 & $\ket{\rm{GHZ}^{d=3}(20^\circ)}_2$ & $2\times2$ & $10^8$ & 15.057 \\
& 2 & $\ket{\rm{GHZ}^{d=3}(25^\circ)}_2$ & $2\times2$ & $10^8$ & 19.363 \\
\label{asymmGHZqt2} & 2 & \bf{$\ket{\rm{GHZ}^{d=3}(29.24^\circ)}_2$ [asym]} & $2\times2$ & $10^9$ & 22.317 \\
\label{symmGHZqt2} & 2 & \bf{$\ket{\rm{GHZ}^{d=3}(35.26^\circ)}_2$ [sym]} & $2\times2$ & $10^9$ & 24.011 \\
& 2 & \bf{$\ket{\rm{GHZ}^{d=3}(35.26^\circ)}_2$ [sym]} & $3\times3$ & $10^7$ & 78.667 \\
& 2 & \bf{$\ket{\rm{GHZ}^{d=3}(35.26^\circ)}_2$ [sym]} & $4\times4$ & $10^7$ & 98.229 \\
& 2 & $\ket{\rm{GHZ}^{d=3}(40^\circ)}_2$ & $2\times2$ & $10^8$ & 22.980 \\
& 2 & $\ket{\rm{GHZ}^{d=3}(45^\circ)}_2$ & $2\times2$ & $10^8$ & 19.763 \\
& 2 & $\ket{\rm{GHZ}^{d=3}(50^\circ)}_2$ & $2\times2$ & $10^8$ & 15.054 \\
& 2 & $\ket{\rm{GHZ}^{d=3}(55^\circ)}_2$ & $2\times2$ & $10^8$ & 10.153 \\
& 2 & $\ket{\rm{GHZ}^{d=3}(60^\circ)}_2$ & $2\times2$ & $10^8$ & 6.329 \\
& 2 & $\ket{\rm{GHZ}^{d=3}(65^\circ)}_2$ & $2\times2$ & $10^8$ & 3.638 \\
& 2 & $\ket{\rm{GHZ}^{d=3}(70^\circ)}_2$ & $2\times2$ & $10^8$ & 1.818 \\
& 2 & $\ket{\rm{GHZ}^{d=3}(75^\circ)}_2$ & $2\times2$ & $10^8$ & 0.714 \\
& 2 & $\ket{\rm{GHZ}^{d=3}(80^\circ)}_2$ & $2\times2$ & $10^8$ & 0.174 \\
& 2 & $\ket{\rm{GHZ}^{d=3}(85^\circ)}_2$ & $2\times2$ & $10^8$ & 0.012 \\
& 2 & $\ket{\rm{GHZ}^{d=3}(90^\circ)}_2$ & $2\times2$ & $10^8$ & 0.000 \\
\cline{2-6} 
& 3 & $\ket{\rm{GHZ}^{d=3}_3(0^\circ)}$	    & $2\times2\times2$ & $10^8$ & 53.360 \\\cline{3-6}
& 3 & $\ket{\rm{GHZ}^{d=3}_3(35.26^\circ)}$ & $2\times2\times2$ & $10^8$ & 82.720 \\\cline{3-6} %
& 3 & $\ket{\rm{A}^-}_3$  & $2\times2\times2$ & $10^8$ & 72.328 \\\cline{3-6} %
& 3 & $\ket{\rm{Q}^1_3}$  & $2\times2\times2$ & $10^8$ & 31.371 \\\cline{3-6} %
& 3 & $\ket{\rm{Q}^2_3}$  & $2\times2\times2$ & $10^8$ & 48.506 \\\cline{3-6} %
& 3 & $\ket{\rm{Q}^3_3}$  & $2\times2\times2$ & $10^8$ & 48.564 \\ %
\hline \hline 
\end{longtable}
\end{footnotesize}

\section{Closing remarks}

In this paper we employed linear programming as a useful tool to analyze the nonclassical properties of quantum states. We checked how many randomly generated sets of observables allow for violation of local realism. Most of the conclusions were presented in the previous sections. Here we want to stress that the overall message of the obtained results is that either for many particles or many measurement settings we observe a conflict with local realism for almost any choice of observables (the probability of violation is greater than $99\%$) for typical families of quantum states.

Concerning the nonclassicality of two qutrits, our results are compatible with those presented in \cite{PhysRevA.92.030101}, that is, maximally entangled and maximally nonclassical states coincide.
It is worth mentioning that, in addition, we addressed the apparently paradoxical result obtained in \cite{VERT-LASK-WIES}.
It amounts to the observation, that the products of $k$-qubit GHZ states and $(N-k)$ pure single qubit states are more nonclassical than the $N$ qubit GHZ state, if we employ the robustness of correlations against white noise admixture as a measure of nonclassicality. Our numerical method shows that the probability of violation of local realism for such product states (for $N = 3,4,5$ and $k = 1,..., N$) is the same as for $k$-qubit GHZ state and thus strictly smaller than for the $N$ qubit GHZ state.
This suggests that resistance against noise, although relevant, is not a good quantifier of nonclassicality.

\section{Acknowledgements}

A.R., J.G., and W.L. are supported by NCN Grant No. 2014/14/M/ST2/00818.
F.P. thanks the financial support from CAPES, CNPq, and FACEPE.
T.V. is supported by the Hungarian National Research Fund OTKA (Grant No. K111734).

\appendix
	
\section{Proof}

\label{proofpartialent}	
The following observation is proven below: The probability of violation $\Pv$ for any two-qubit state and two binary-outcome measurements cannot be greater than $2(\pi-3)$.

In order to prove it, first recall that $\Pv=2(\pi-3)$ for the $\ket{\rm{GHZ_2}}$ state using two binary-outcome settings per party~\cite{PhysRevLett.104.050401}. Let
\begin{equation}
\label{partialent}
\ket{\rm{GHZ}(\alpha)}_2=\sin \alpha \ket{00}+\cos \alpha \ket{11}
\end{equation}
stand for the two-qubit pure partially entangled state with $\alpha\in\{0,\pi/4\}$ written in the Schmidt bases. Clearly, $\alpha=\pi/4$ recovers the two-qubit maximally entangled state. Let $\Pv(\alpha)$ denote the probability of violation corresponding to the state $\ket{\rm{GHZ}(\alpha)}_2$. Note that mixed two-qubit states cannot provide higher probability of violation, therefore we can restrict our attention to the probabilities $\Pv(\alpha)$.

Firstly we prove the following lemma: if the CHSH inequality is violated using a state $\ket{\rm{GHZ}(\alpha)}_2$ and some projective measurements, at least the same violation occurs with the maximally entangled state $\ket{\rm{GHZ}}_2$ using the same measurements.
\begin{proof}
Let us write the measurement observables $A$ and $B$ as
\begin{align}
A &= a_x\sigma_x + a_y\sigma_y + a_z\sigma_z,\nonumber\\
B &= b_x\sigma_x + b_y\sigma_y + b_z\sigma_z,
\end{align}
where $\sigma_{x,y,z}$ denote Pauli matrices, and the coefficients of Alice measurements $a_x, a_y, a_z$ square to 1 (and similarly for Bob). Then we have the joint correlator
\begin{equation}
\label{AB}
\langle AB\rangle = a_zb_z + \sin2\alpha(a_xb_x -a_yb_y).
\end{equation}
On the other hand, the CHSH expression reads
\begin{equation}
\label{CHSH}
\textrm{CHSH} = \langle A_1B_1\rangle + \langle A_1B_2\rangle + \langle A_2B_1\rangle - \langle A_2B_2\rangle.
\end{equation}
Using formula~(\ref{AB}), we get
\begin{equation}
\label{CHSH_C}
\textrm{CHSH}(\alpha) = C_z + \sin(2\alpha)(C_x - C_y),
\end{equation}
where $\alpha\in\{0,\pi/4\}$ and
\begin{equation}
C_i = a_{1i}b_{1i} + a_{1i}b_{2i} +a_{2i}b_{1i} -a_{2i}b_{2i},
\end{equation}
where $i$ can take $x$, $y$, and $z$. Notice that $C_z\le2$, therefore a CHSH value greater than 2 in equation~(\ref{CHSH_C}) implies that $C_x-C_y>0$. This in turn implies that in case of violation of the CHSH inequality (that is $\rm{CHSH}>2$), we have $\rm{CHSH}(\pi/4)\ge \rm{CHSH}(\alpha)$ for all $\alpha\in\{0,\pi/4\}$.
\end{proof}

Given the above lemma, it is not difficult to see that $\Pv(\alpha)\le\Pv=2(\pi-3)$ for all $\alpha\in\{0,\pi/4\}$.
Indeed, notice that the classically attainable region of the 2-setting 2-outcome scenario is completely characterized by eight different versions of the CHSH expressions (see e.g.~\cite{PhysRevLett.104.050401}). However, after suitable relabeling of the inputs and flipping of the outcomes they all end up in the standard CHSH defined by equation \ref{CHSH}. Hence, violation of any of the versions of the CHSH inequality using a partially entangled state~(\ref{partialent}) along with some projective measurements entails at least the same violation of this version using the maximally entangled state and the same measurements. This implies the relation $p_V(\alpha) \le p_V = 2(\pi-3)$  we set out to prove.
\end{document}